\documentclass[journal=apchd5,manuscript=article]{achemso}
\usepackage{array}
\usepackage{comment}
\usepackage[version=3]{mhchem} 

\title{On-Chip Glucose Sensing at Terahertz Frequencies}

\author{Mohsen Haghighat}
\affiliation{Department of Electrical and Computer Engineering, University of Victoria, Victoria, British Columbia V8P 5C2, Canada}
\altaffiliation{Centre for Advanced Materials and Related Technology (CAMTEC), University of Victoria, Victoria, British Columbia V8W 2Y2, Canada}

\author{Thomas Darcie}
\affiliation{Department of Electrical and Computer Engineering, University of Victoria, Victoria, British Columbia V8P 5C2, Canada}

\author{Levi Smith}
\affiliation{Department of Electrical and Computer Engineering, University of Victoria, Victoria, British Columbia V8P 5C2, Canada}
\altaffiliation{Centre for Advanced Materials and Related Technology (CAMTEC), University of Victoria, Victoria, British Columbia V8W 2Y2, Canada}

\email{levismith@uvic.ca}
\affiliation{Department of Electrical and Computer Engineering, University of Victoria, Victoria, British Columbia V8P 5C2, Canada}

\abbreviations{Coplanar Strip: CPS, Terahertz: THz, Time Domain Spectroscopy: TDS, Frequency Domain: FD, Coplanar Waveguide: CPW, Microstrip Line: MSL}
\keywords{D-glucose, Guided Wave, On-Chip Sensing, Coplanar Strip (CPS), Terahertz (THz), Thin Membrane, Silicon Nitride}

\begin{document}

\begin{abstract}
This paper demonstrates an on-chip anhydrous D-glucose sensor based using a coplanar stripline (CPS) on a thin (1 \textmu m) silicon nitride membrane at terahertz (THz) frequencies. A thin layer ($\approx$ 10 \textmu m) of D-glucose was placed in close proximity to the CPS and the transmission response was measured using a modified THz-TDS setup. The D-glucose introduces frequency-dependent changes to the effective permittivity of the CPS resulting in a modified spectral response at the receiver. Measurement results show absorption signatures at 1.42 THz and 2.07 THz corresponding to the first two significant resonances beyond 1 THz for D-glucose allowing for label-free detection. The frequency-dependent attenuation coefficient was estimated by simulation for several D-glucose layer thicknesses using a modified Lorentz model. Measurement results align with simulations and other literature that use free-space THz radiation. This work verifies on-chip THz sensing of D-glucose and presents a pathway towards on-chip sensing of other materials at THz frequencies.
\end{abstract}

\section{Introduction}

Glucose detection is important across various industries, including the food, beverage, and medical sectors \cite{Hassan2021_Glucose_Recent_advances_MDPI_Sensors}. The escalating consumption of sugar in diets has been linked to chronic health issues, notably cardiovascular diseases, diabetes, and obesity \cite{Mozaffarian2016_diabetes_obesity_circulation,ventura2011_obesity_diabetes_with_sugar}. Given the importance of these health issues, accurate detection and monitoring of glucose both in the human body and food become imperative. In response to this need, researchers have been developing glucose-sensing technologies for blood glucose monitoring in the last couple of decades \cite{Oliver2009_Review_Glucose_Sensors_diabetic_medicine, Filipo2023_Review_Non-Invasive_Glucose_Sensing_MDPI_Sensors}. Also, industries such as food, beverage, and fermentation manufacturing are required to shrink and control sugar levels while ensuring the quality and safety of their products \cite{Hassan2021_Glucose_Recent_advances_MDPI_Sensors}. This necessitates continuous monitoring and measurement of various samples, including sugars, throughout the manufacturing process and in the final product.  On the other hand, selective detection of glucose is essential for investigation in the food and beverage industry because natural glucose should not be entirely replaced with artificial sweeteners since they have other health side-effects such as sugar craving and mood disorders \cite{Choi_2022_natural_artificial_sweetener}. The necessity to either control excessive sugar levels or detect natural sugars from artificial sweeteners motivates researchers to explore advanced sugar-sensing methodologies. D-glucose is one of the main sugars which exists in food and the common naturally occurring simple sugar \cite{D-glucose}. In the context of this paper, the term `glucose' exclusively denotes anhydrous D-glucose \cite{Liu2006_Dehydrated_Glucose_1.43_2.1_CPL}. 

The pursuit of glucose sensing methods aligns with the broader objective of promoting health and well-being, either glucose detection in food, or blood glucose monitoring. Glucose sensing methods have been reviewed in \cite{Oliver2009_Review_Glucose_Sensors_diabetic_medicine, Filipo2023_Review_Non-Invasive_Glucose_Sensing_MDPI_Sensors} from traditional enzymatic and electrochemical methods to optical and spectroscopic techniques. Regarding in-body glucose sensing, the category of optical sensors is favorable for their non-invasive nature and potential for continuous monitoring. These methods include fluorescence, Raman spectroscopy, near-infrared (NIR), mid-infrared (MIR), and far-infrared (FIR) spectroscopy \cite{Filipo2023_Review_Non-Invasive_Glucose_Sensing_MDPI_Sensors}. However, optical methods face challenges due to scattering \cite{Oliver2009_Review_Glucose_Sensors_diabetic_medicine,Pickwell_2006_THz_bio_applications} with reduced scattering coefficient of 30 - 50 cm$^{-1}$ \cite{Jacques_2013_Optical_properties_of_tissue_scattering}, contributing to signal-to-noise ratio (SNR) reduction \cite{Truong2022_Influence_of_SNR_scatter_FOPT_}, relatively weak glucose absorption peaks especially in NIR range \cite{Kottmann2012_Glucose_epidermis_mid_IR_weak}, and interference from surrounding light \cite{Hassan2021_Glucose_Recent_advances_MDPI_Sensors}. On the other hand, THz waves have less scattering compared to IR and optical beams \cite{Pickwell_2006_THz_bio_applications,Truong2022_Influence_of_SNR_scatter_FOPT_} with estimated reduced scattering coefficient up to 1 cm$^{-1}$ based on analysis in \cite{Jacques_2013_Optical_properties_of_tissue_scattering}, and have strong glucose absorption peaks \cite{Lee_2015_Highly_gluceose_sensing_nano_antennas_nsrep,Liu2006_Dehydrated_Glucose_1.43_2.1_CPL} which enables easier glucose detection with less noise. Moreover, THz waves penetrate into human tissue up to 300 \textmu m \cite{Vilagosh_2019_THz_penetration_tissue_0.2mm}, potentially reaching the dermis layer where blood and glucose are present whereas the MIR range only penetrates up to 100 \textmu m, which corresponds to the thickness of the epidermis layer that lacks blood \cite{Kottmann2012_Glucose_epidermis_mid_IR_weak}. However, glucose can still be detected in the MIR and THz range within the interstitial fluid (ISF) of the epidermis, where a slight concentration of glucose is present but blood capillaries are not reached at this depth \cite{Kottmann2012_Glucose_epidermis_mid_IR_weak}. Absorption coefficients extracted from THz time-domain spectroscopy (THz-TDS) measurements have shown a high sensitivity to the glucose level in blood samples \cite{Chen2018Quantify_glucose_THz_TDS}. Additionally, THz-TDS combined with weak value amplification were demonstrated to be sensitive to small changes in concentrations of glucose both in solid and liquid samples \cite{Lu2023_WVA_THz_TDS_ACS_Photonics}. However, challenges remain in terms of miniaturization and cost-effectiveness of THz sensors \cite{Filipo2023_Review_Non-Invasive_Glucose_Sensing_MDPI_Sensors}, which can be addressed by on-chip THz sensors, that are the focus of this research.


Glucose absorbs electromagnetic radiation at certain frequencies in the THz band. The first peak of absorption spectra for glucose beyond 1 THz occurs in the range of 1.40 - 1.44 THz \cite{Lee_2015_Highly_gluceose_sensing_nano_antennas_nsrep, Song_2018_nsrep_absorption_1.42THz_solid_glucose,Jaber2024_Jean-michel_Menard_Nature_Communicaton,Liu2006_Dehydrated_Glucose_1.43_2.1_CPL,LEE2023_Imoroved_Modified_Lorentz_1.435_Spectrochimica, Huang_2022_glucose_spectrum_1.44THz}. In literature, the majority of glucose sensing methods at THz frequencies involve directing a free-space THz beam towards a thin glucose sample \cite{Liu2006_Dehydrated_Glucose_1.43_2.1_CPL}, a metasurface \cite{Beruete_2019_AOM_Review_of_THz_sensing_metasurfaces, Yang_2021_glucose_concentration_metasurface_400GHz, Serita_2019_MDPI_phptonic_THz_microfluidic_glucose_400GHz_meta, Jaber2024_Jean-michel_Menard_Nature_Communicaton, LEE2023_Imoroved_Modified_Lorentz_1.435_Spectrochimica}, or nano-antennas (i.e., also a form of metasurface) \cite{Lee_2015_Highly_gluceose_sensing_nano_antennas_nsrep} coated by sugar (glucose) then detecting the transmitted wave. Using metasurfaces results in strong subwavelength field concentration 
at a resonance frequency, making them sensitive to small features and slight variations in the nearby environment, thereby enhancing their overall sensitivity \cite{Beruete_2019_AOM_Review_of_THz_sensing_metasurfaces}. In \cite{Serita_2019_MDPI_phptonic_THz_microfluidic_glucose_400GHz_meta, Yang_2021_glucose_concentration_metasurface_400GHz} authors designed metasurface unit cells that have resonance frequencies below 0.5 THz and then measured the shift in resonance frequency in the presence of glucose or other materials. Others \cite{Lee_2015_Highly_gluceose_sensing_nano_antennas_nsrep, Jaber2024_Jean-michel_Menard_Nature_Communicaton} have designed the metasurface unit cell so that it targets the first THz-band absorption frequency of glucose near 1.4 THz. 
In \cite{Lee_2015_Highly_gluceose_sensing_nano_antennas_nsrep} the authors proposed a metasurface for THz transmission
by nano-slot-antennas which increases the molecular absorption cross-sections, thereby enabling
the detection of sugar molecules at low concentrations. Table \ref{tab:glucose_sensing} provides an overview of various THz sensing methods for sugar detection including key details such as the reference source, resonance frequency in THz, frequency range covered by the sensor, structural design, mechanism of wave transmission, spectral detection method employed, and the type of sugar targeted in the respective study. Our work, highlighted as `This Work' in Table \ref{tab:glucose_sensing}, proposes on-chip glucose sensing operating in the broad frequency range of 0.1 - 2.2 THz, employing guided waves which as previously mentioned, is the key difference from the other works.

The aforementioned studies use free-space THz radiation for sensing which offers simplicity and flexibility in accommodating various sample geometries and configurations; however, free-space methods suffer from path loss and require bulky THz optical components \cite{Zeng2022_THz_PM_free-space_guided-wave}. Alternatively, guided-wave methods are attractive because they are compact and do not require the alignment of a THz beam. Also, a guided-wave sensor can offer lower channel loss \cite{Zeng2022_THz_PM_free-space_guided-wave} and increased sensitivity by confining the electromagnetic field within the sample region making it suitable for on-chip integrated sensors in various applications \cite{Singh_2020_SSPP_Sensor}.

There are several studies that use guided-waves for sensing at THz frequencies. In \cite{Calello_2021_On_chip_sensing_high_loss_liquids_CPW_TTST} the authors reported high-loss liquid sensing including water using a coplanar waveguide (CPW) up to 1.1 THz where the signal decays below the noise floor. A similar approach is taken with microstrip lines (MSL) in \cite{Ohkubo2006_Microstrip_THz_liquid_sensing} for liquid sensing (i.e., water), and authors mentioned that the experimental results for frequencies greater than 1 THz are close to the noise floor, although spectral data is reported up to 1.5 THz. This is mainly due to the high loss of MSL configuration at THz frequencies and also partly because of the thickness of the waveguide cover layer \cite{Ohkubo2006_Microstrip_THz_liquid_sensing}.

\begin{figure}
    \centering
    \includegraphics[width=0.6\linewidth]{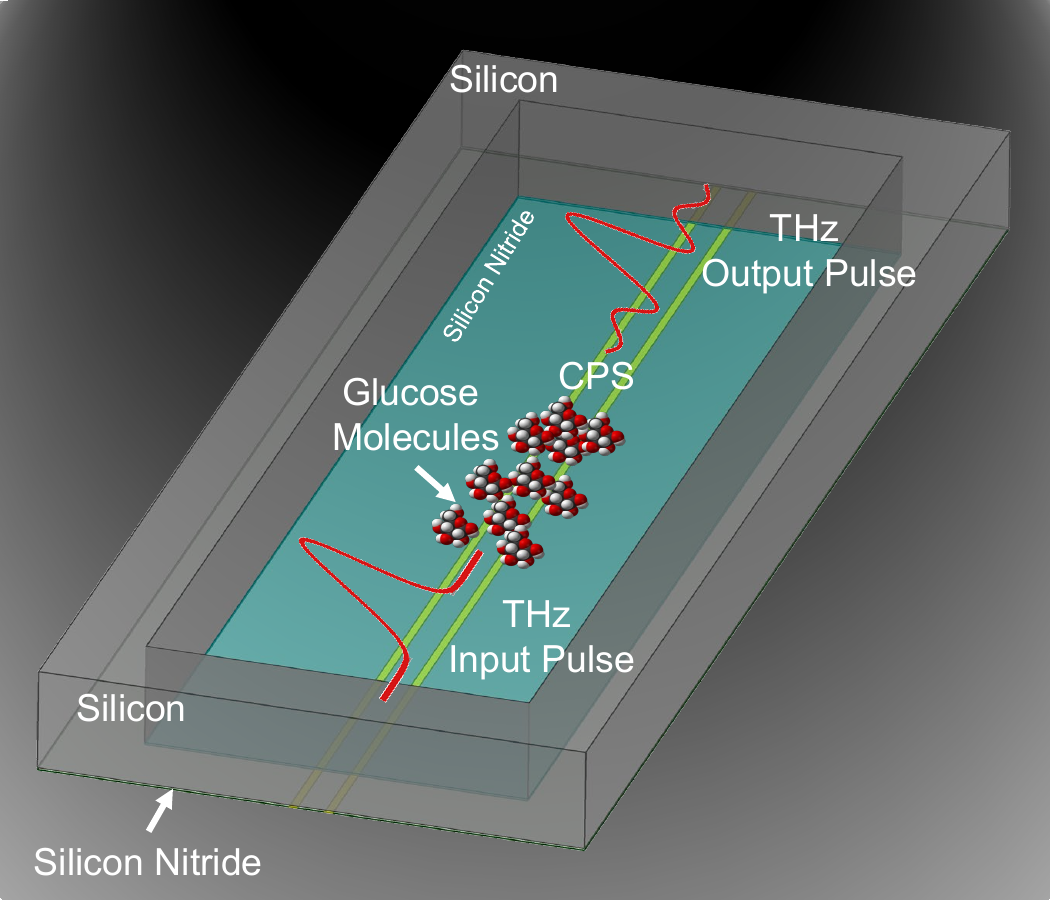}
    \caption{Illustration of glucose sensing by CPS waveguide on a thin \ce{Si3N4} membrane}
    \label{fig:glc_concept_illustration}
\end{figure}

This work explores a unique approach for on-chip glucose sensing using guided waves via coplanar stripline (CPS) on a thin silicon nitride (Si$_3$N$_4$) substrate which presents a platform that mitigates the loss and limitations of standard THz waveguide designs \cite{cheng1994terahertz, Levi_Smith_CPS_on_Si3N4_1st}. Also, the CPS on a thin membrane enables the interaction of the electromagnetic field with materials (such as glucose) beneath the membrane (see Fig. \ref{fig:glc_concept_illustration}). Our experiments show the prominent THz-band absorption frequency of 1.42 THz measured with the fabricated sensor, which aligns with the absorption frequency reported in Table \ref{tab:glucose_sensing}. We note that label-free sensing involves the detection of the multiple characteristic absorption signatures which occur at different frequencies and the measurement of a broad spectrum facilitates this capability. To confirm label-free sensing, we simultaneously resolve the glucose absorption features at 1.42 THz and 2.07 THz.   \cite{Liu2006_Dehydrated_Glucose_1.43_2.1_CPL,Song_2018_nsrep_absorption_1.42THz_solid_glucose}.

\begin{table}
    \centering
    \small  \small  \small
    \begin{tabular}{m{1.5cm}m{2cm}m{1.5cm} m{2cm}m{2cm}m{2cm}m{2cm}}
        \hline
        \textbf{Ref.} & \textbf{Resonance Freq. (THz)} & \textbf{Freq. Range (THz)} & \textbf{Structure} & \textbf{Wave transmission} & \textbf{Spectral Detection Method} & \textbf{Sugar type} \\
        \hline
        \hline        
        [\cite{Liu2006_Dehydrated_Glucose_1.43_2.1_CPL}] & 1.43, 2.08 & 0.3-3.0 & NA (direct exposure) & Free-Space &Absorption frequency & D-glucose  \\
        \hline 
        [\cite{Lee_2015_Highly_gluceose_sensing_nano_antennas_nsrep}] & 1.4, 1.7 & 0.5-2.5 & Nano-slot antennas & Free-Space & Absorption frequency  & D-glucose, Fructose, Sucrose \\
        \hline
        [\cite{Jaber2024_Jean-michel_Menard_Nature_Communicaton}] & 1.43 & 1.2-1.6 & Metasurface & Free-Space &Absorption frequency & D-glucose  \\
        \hline            [\cite{LEE2023_Imoroved_Modified_Lorentz_1.435_Spectrochimica}] & 1.435 & 1.2-1.6 & Metasurface & Free-Space &Absorption frequency & D-glucose  \\
        \hline

        [\cite{ Yang_2021_glucose_concentration_metasurface_400GHz}] & 0.26 & 0.2-0.5 & Metasurface & Free-Space & Resonance shift & D-glucose \\       
        \hline       
        [\cite{Serita_2019_MDPI_phptonic_THz_microfluidic_glucose_400GHz_meta}] & 0.32 & 0.2-0.5 & Metasurface & Free-Space & Resonance shift & D-glucose \\       
        \hline
        [\cite{Hu_2016_metamaterial_absorber_glucose_ethanol}] & 0.6 & 0.3-1.2 & Meta-material absorber & Free-Space & Resonance shift & D-glucose \\       
        \hline        
        This Work & 1.42, 2.07 & 0.1-2.2 & CPS & Guided- Wave & Absorption frequency & D-glucose \\
        \hline            
    \end{tabular}
    \caption{Summary of experimental sugar sensing methods based on THz spectroscopy}
    \label{tab:glucose_sensing}
\end{table}

\section{Design and Methods}

\begin{figure}
    \centering
    \includegraphics[width=0.75\linewidth]{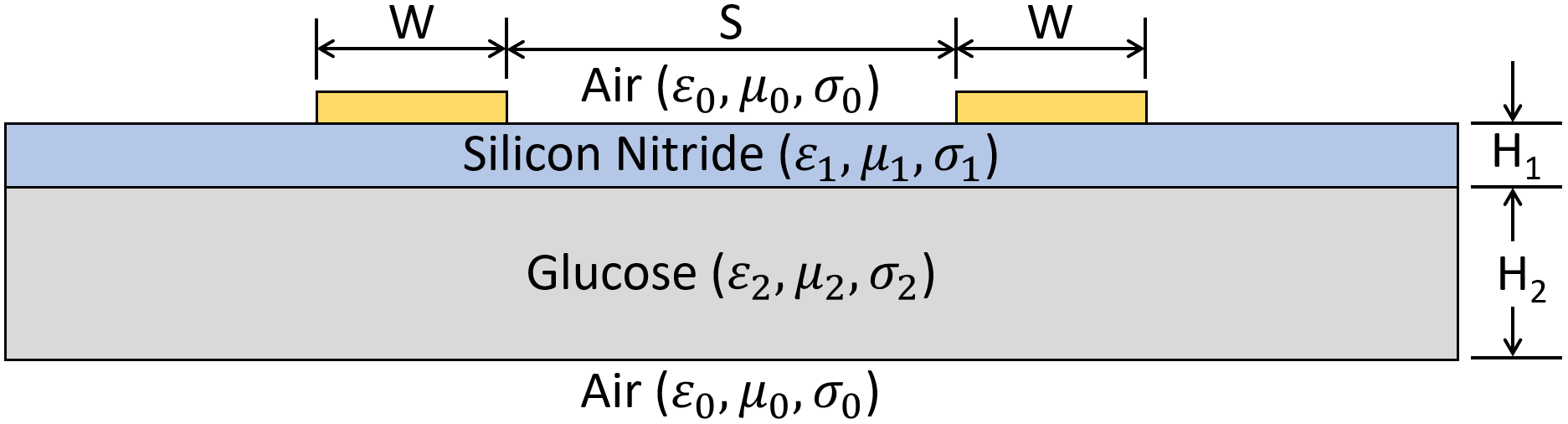}
    \caption{CPS sensor cross section}
    \label{fig:xsect}
\end{figure}

The sensor presented in this work operates on the principle of frequency-dependent dielectric loss from a nearby material (i.e., glucose). Figure \ref{fig:xsect} illustrates a cross-section of the sensor geometry which is a CPS transmission line ($S$ = 130 \textmu m and $W$ = 30 \textmu m) on a suspended thin \ce{Si3N4} membrane ($H_1$ = 1 \textmu m) which is in contact with a thin glucose layer ($H_2$ $\approx$ 10 \textmu m). The effective permittivity of the propagating wave, \(\varepsilon_{\text{eff}}\), depends on the individual material parameters ($\varepsilon_{\text{0,1,2}}, ~\mu_{\text{0,1,2}},~\text{and}~\sigma_{\text{0,1,2}}$) and the dimensions of the transmission line (S, W, and H$_{1,2}$). Here, of primary concern, is detecting frequency-dependent variations in glucose permittivity, \(\varepsilon_{\text{2}} (\omega)\), which can be achieved when the \ce{Si3N4} layer is thin H$_1 \approx$ 1 \textmu m.

\begin{figure}
    \centering
    \includegraphics[width=1\linewidth]{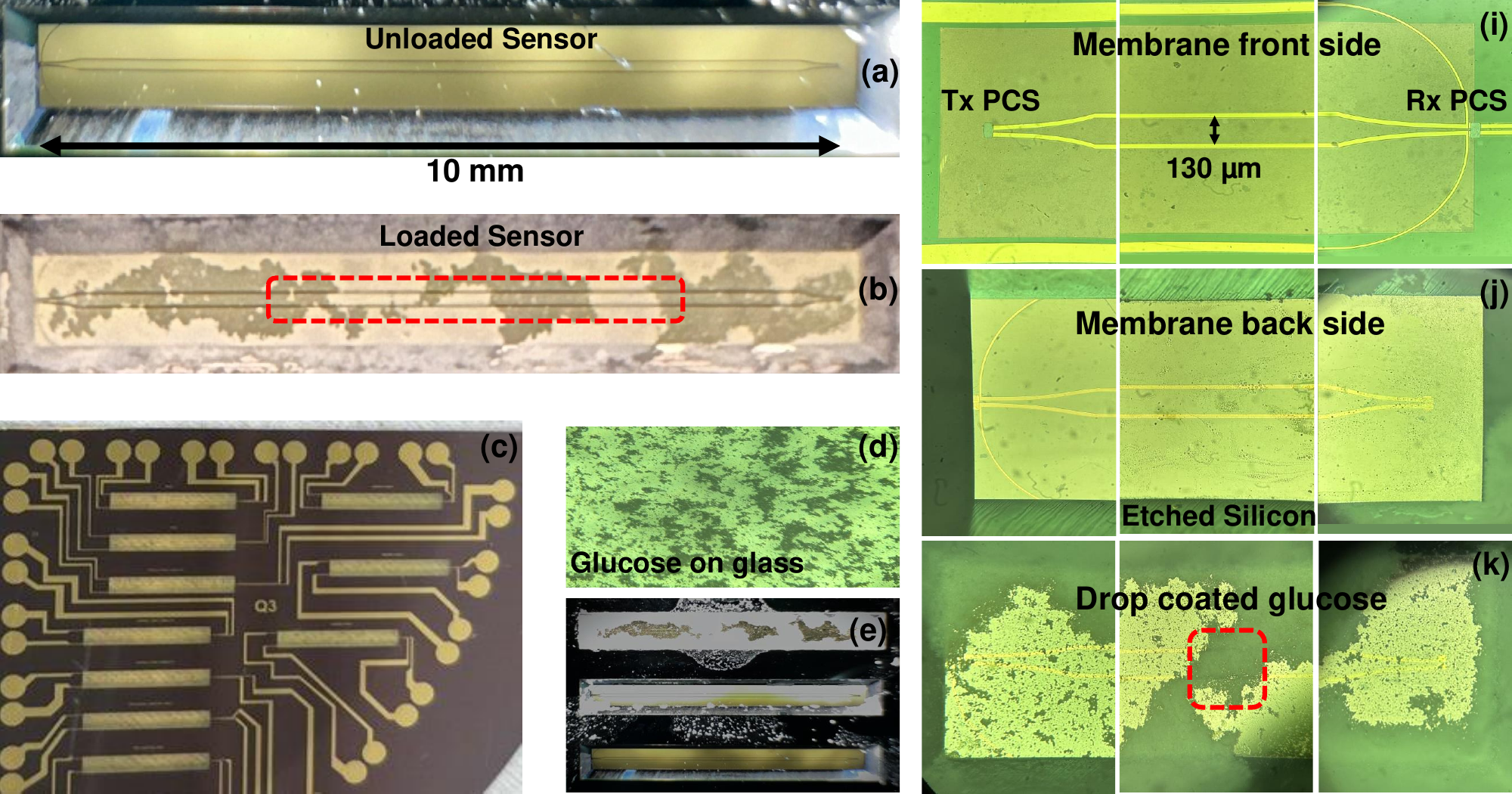}
    \caption{Sensor and membrane pictures. (a) Clean (unloaded) membrane back side.
(b) Dehydrated glucose on the back side membrane (loaded sensor). 
(c) Fabricated wafer.
(d) glucose grains on a glass after 10 minutes of sonication with isopropanol. (e) CPS sensor windows (wafer backside). (i)  Membrane front: CPS side microscope image.
(j) Membrane back: Material side microscope image.
(k) Membrane back image with dried drop-coated glucose with a micropipette}
    \label{fig:Membrane_Glucose_illustration}
\end{figure}

\begin{figure}
    \centering
    \includegraphics[width=0.7\linewidth]{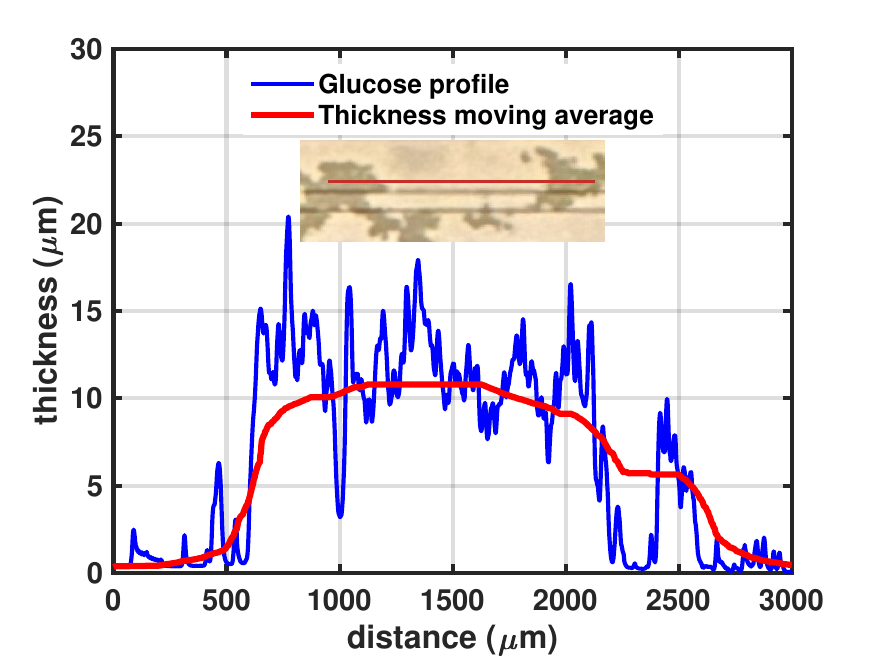}
    \caption{Thickness profile of the glucose layer, and back side of the \ce{Si3N4} membrane with non-uniform drop-coated glucose (inset).}
    \label{fig:Glucose_Profile}
\end{figure}

The CPS sensor structure is fabricated by depositing a 1 \textmu m \ce{Si3N4} layer on a 500 \textmu m silicon substrate. Next, a rectangle of silicon with a dimension of 2000 \textmu m $\times$ 10000 \textmu m is etched under the structure where the sample material will be placed. Metallic conductors are placed using photolithography and metal deposition (Ti/Au: 10 nm/200 nm). We made a mixture of glucose and isopropanol (5 grams: 50 mL) and sonicated that in a beaker to reduce the grain size. Then, we placed drops of the liquid using a micropipette in steps of 10 \textmu L on the backside of the membrane and waited for the isopropanol to evaporate. Figure \ref{fig:Membrane_Glucose_illustration} illustrates microscopic and close-up photos of the sensor from with and without glucose on the backside of the membrane.
The thickness of the resulting glucose layer was measured using a profilometer (Figure \ref{fig:Glucose_Profile}) after placing 60 \textmu L of the mix.  It is important to note that with our current simple drop placement methodology precise control of the glucose distribution was not possible. Future work will address this limitation by the introduction of lithographically-defined microfluidic channels. For this specific experiment, the glucose film thickness is H$_2 \approx 10$ \textmu m, and the glucose interaction length is L $\approx$ 3 mm (Figure \ref{fig:Membrane_Glucose_illustration}b). We note that to avoid absorption saturation, a thin sample layer is desired \cite{Liu2006_Dehydrated_Glucose_1.43_2.1_CPL}.

\section{Modeling and Simulation}

In this study, we incorporate a frequency-dependent permittivity model for glucose into the existing model of a CPS that is depicted in Figure \ref{fig:xsect}. Our objective is to obtain the attenuation coefficient of the CPS for glucose layers of varying thicknesses for frequencies ranging from 0.1 to 2.7 THz. The simulation was performed using ANSYS HFSS 2023 and characterizes the CPS behavior in the presence of glucose at different frequencies and thicknesses. The material properties of the model components include the \ce{Si3N4} substrate with 1 \textmu m thickness and $\varepsilon_r$ = 7.6, $\sigma_{Si_3N_4}=0$, \mbox{tan $\delta_e$ = 0.00526} \cite{Cataldo_Silicon_nitride_properties_2012}, and gold conductors with 200 nm thickness and the conductivity of $\sigma_{Au} = 4.1 \times 10^7$ S/m. The dielectric permittivity of the glucose layer with single resonance at THz frequencies is modeled using the modified Lorentz model presented in (\ref{eqn:dielectric})
\cite{LEE2023_Imoroved_Modified_Lorentz_1.435_Spectrochimica}:

 
\begin{equation}
{\varepsilon}(\omega) = \varepsilon_\infty + \frac{\omega_p^2}{\omega_0^2 - \omega^2 + i\gamma\omega}
\label{eqn:dielectric}
\end{equation}

where $\varepsilon_\infty = 3.231$ is the dielectric coefficient at infinite frequency \cite{LEE2023_Imoroved_Modified_Lorentz_1.435_Spectrochimica}, $\omega_0 = 2\pi(1.435 \, \text{THz}$) is  the resonance frequency, $\omega_p = 2\pi(0.240 \, \text{THz}$) is bulk plasma frequency of the dielectric, and $\gamma = 2\pi(0.416 \, \text{THz}$) represents the damping constant \cite{LEE2023_Imoroved_Modified_Lorentz_1.435_Spectrochimica}.

\begin{figure}
\begin{minipage}[b]{0.45\linewidth}
        \centering
        \includegraphics[width=\linewidth]{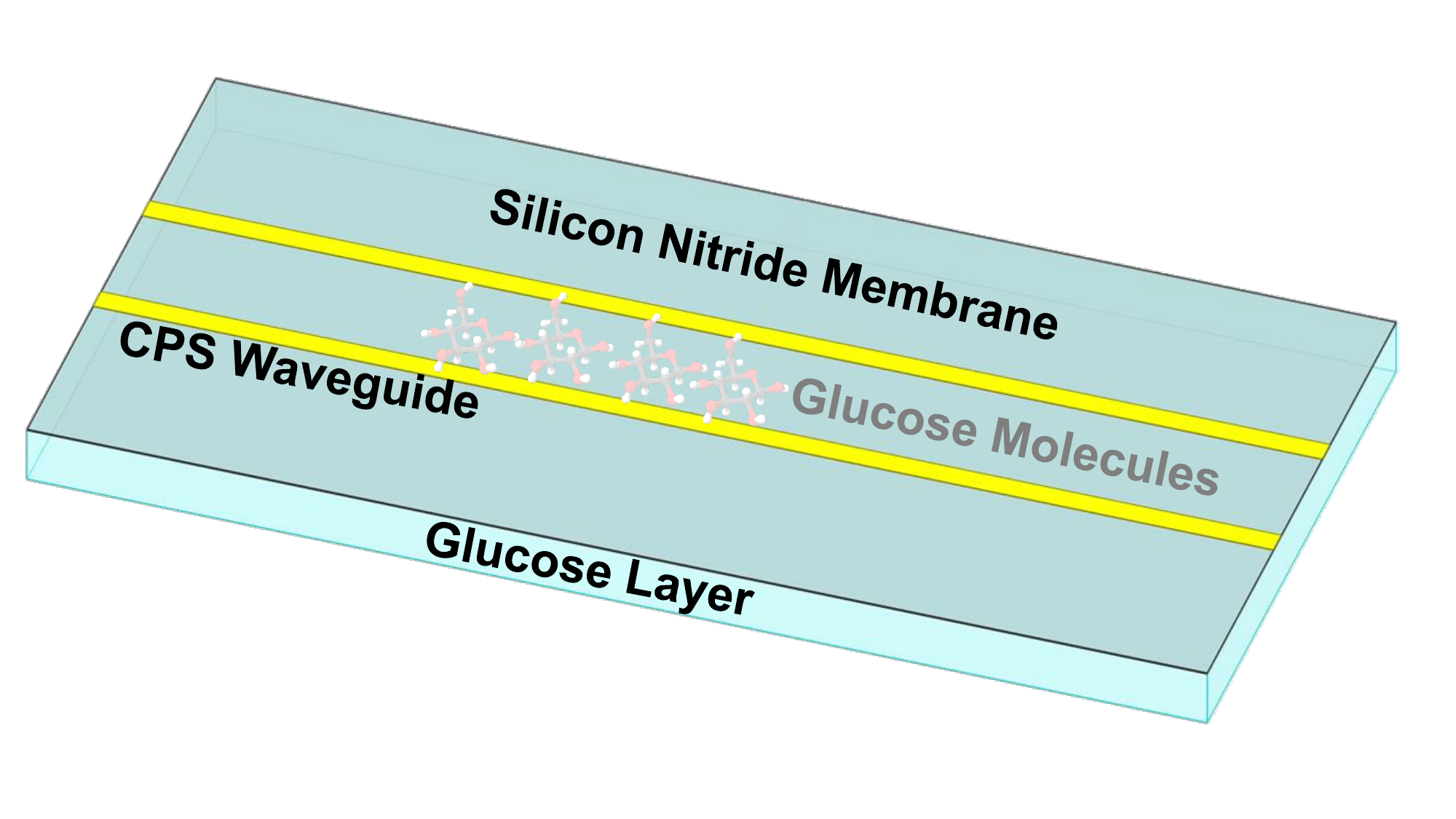}
        \caption*{(a)}
    \end{minipage}
    \begin{minipage}[b]{0.54\linewidth}
        \centering
        \includegraphics[width=1\linewidth]{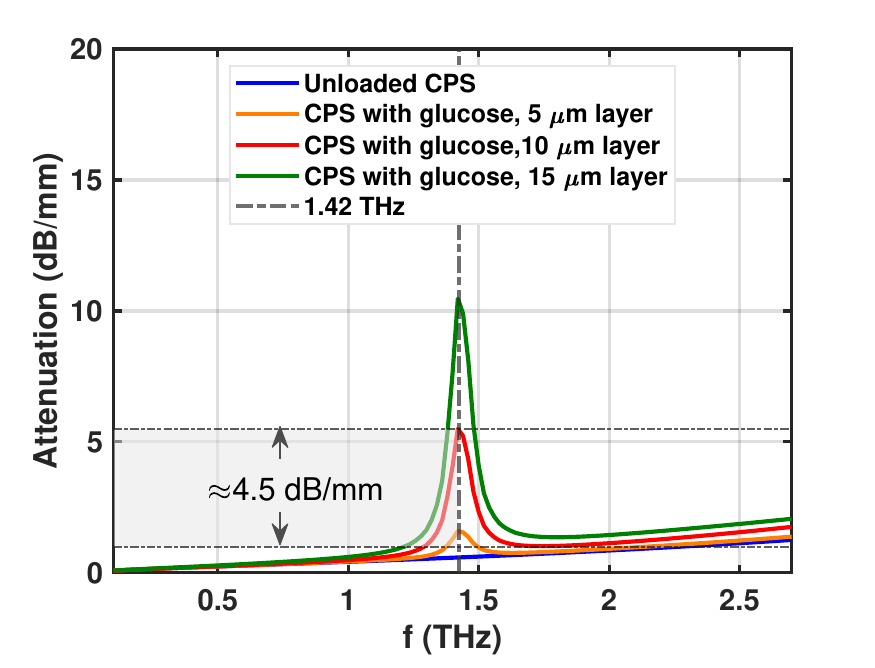}
        \caption*{(b)}
    \end{minipage}
    
    \caption{(a) CPS on the thin membrane with glucose layer. (b) Attenuation of CPS with modified Lorentz model for glucose layer with different thicknesses under the sensor's thin membrane.}
    \label{fig:3D_Model_Attenuation_with_glucose}
\end{figure}

Next, we simulated the attenuation coefficient for the CPS illustrated in Figure \ref{fig:xsect} using (\ref{eqn:dielectric}) to represent $\varepsilon_{\text{2}}$ for several different glucose thicknesses (H$_2$ = 5, 10, 15 \textmu m). The results of this simulation are plotted in Figure \ref{fig:3D_Model_Attenuation_with_glucose}(b) along with a reference without any glucose which has an attenuation of $\approx$ 0.6 dB/mm at 1.4 THz. As previously mentioned, the average thickness of glucose is $\approx$ 10 \textmu m based on the average of the profile plot in Figure \ref{fig:Glucose_Profile} measured by a profilometer along the top CPS line depicted on the inset. According to Figure \ref{fig:3D_Model_Attenuation_with_glucose}(b) this thickness corresponds to an attenuation depth of $\approx$ 4.5 dB/mm (i.e. the height of the absorption peak at 1.42 THz).

\section{Results and Discussion}

\begin{figure}
    \centering
    \includegraphics[width=4in]{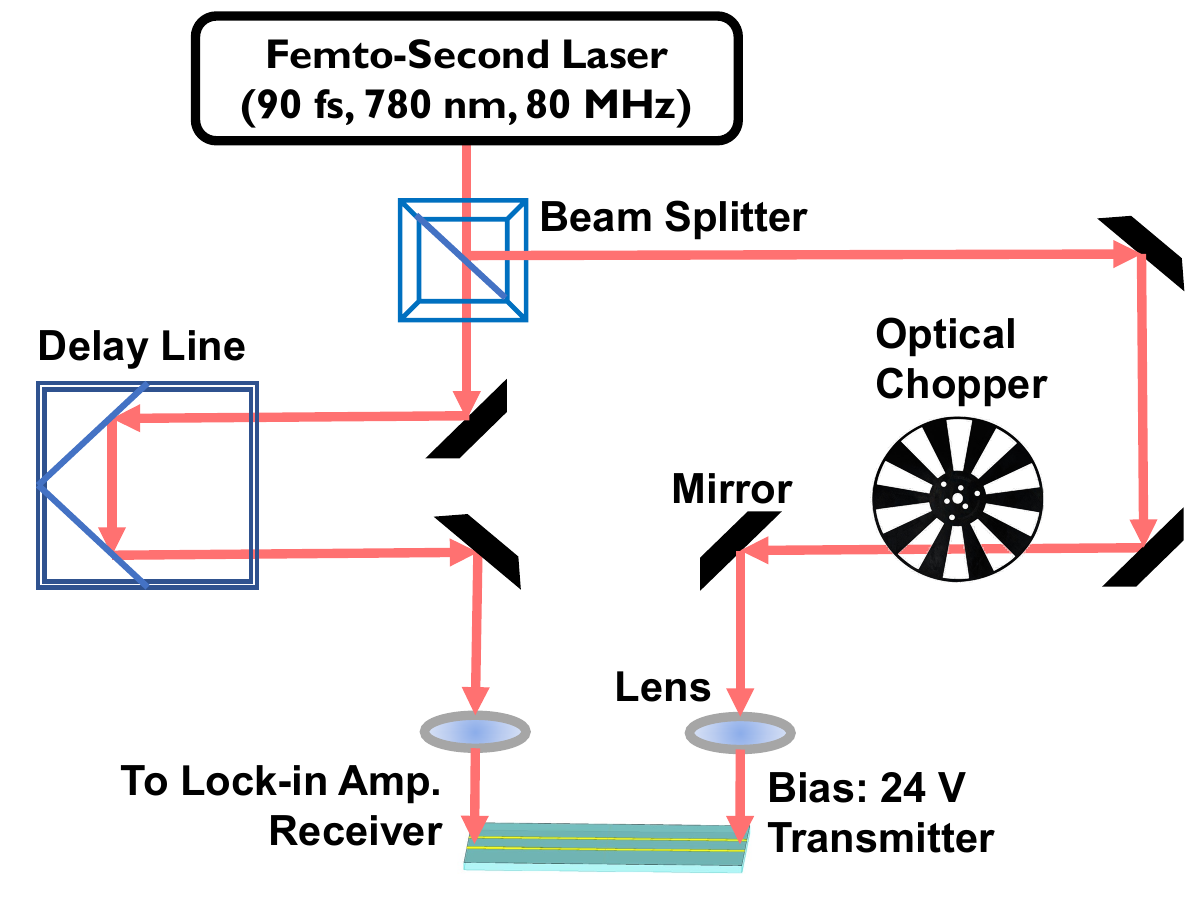}
    \caption{Measurement setup diagram for glucose sensing.}
    \label{fig:Glucose_CPS_Measurement_Setup}
\end{figure}

The experimental measurements were carried out using a modified THz-TDS setup  that is in Figure \ref{fig:Glucose_CPS_Measurement_Setup} \cite{Levi_Smith_CPS_on_Si3N4_1st}. A femtosecond pulsed laser with a wavelength of 780 nm and an average output power of 20 mW was split and directed towards thin-film low-temperature grown gallium arsenide (LT-GaAs) photo-conductive switches (PCS) placed on both sides of the tapered CPS to generate and detect a broadband THz pulse signal \cite{Levi_Smith_CPS_on_Si3N4_1st}. The fabrication of the LT-GaAs PCS devices is detailed in \cite{Rios2015_bowtie_PCA, Haghighat2024_CPS_SSPP_SREP}. The transmitted signal was reconstructed by photo-conductive sampling of the output of the THz pulse, using the mechanical delay line and measuring the receiver current using a lock-in amplifier, similar to conventional THz-TDS techniques \cite{Jepsen2011_THzSpec}.

\begin{figure}
    \centering    
    \includegraphics[width=\linewidth]{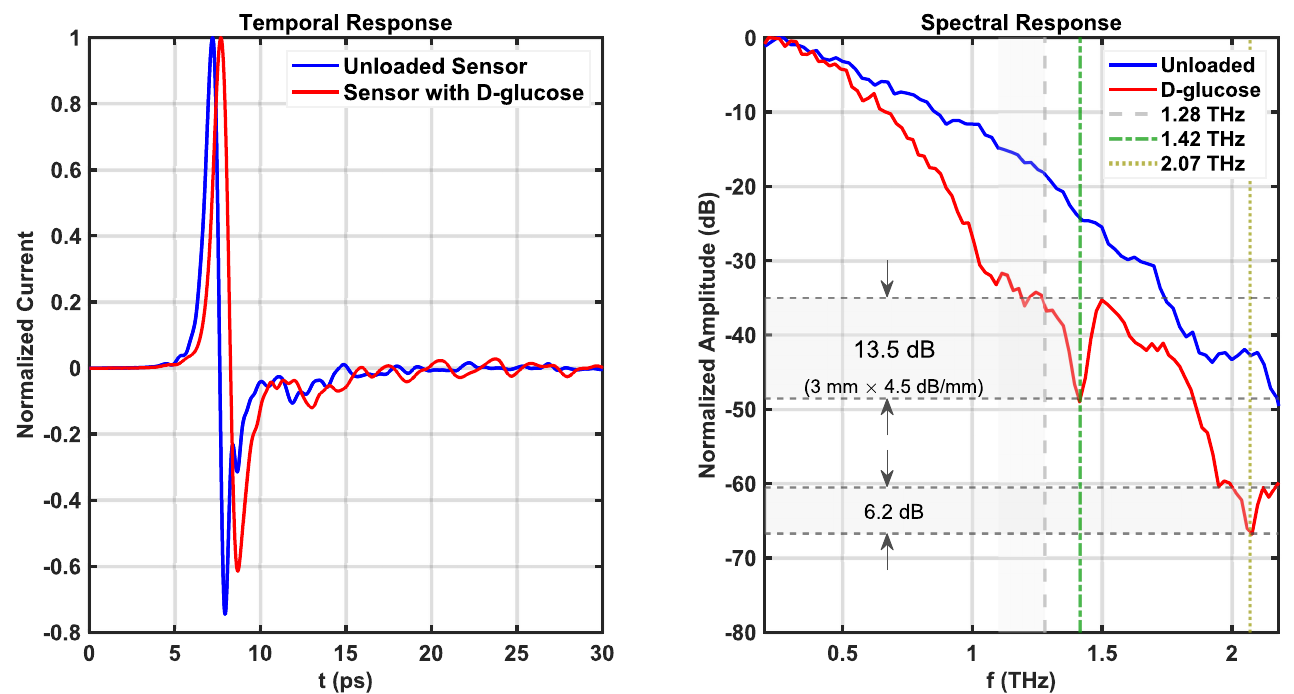}
    \caption{Measurement results of the proposed CPS-based glucose sensor}    \label{fig:Glucose_sensing_CPS_measurement_results}
\end{figure}

The measurement results of unloaded (no glucose) and loaded sensor (with $\approx$ 10 \textmu m glucose layer) are shown in Figure \ref{fig:Glucose_sensing_CPS_measurement_results}. Note that the spectral amplitude decay with increasing frequency in the experimental results is an expected consequence of applying a THz finite duration pulse as an input signal rather than an ideal THz impulse with a flat spectral response \cite{Levi_Smith_CPS_on_Si3N4_1st}.
From the simulations (Figure \ref{fig:3D_Model_Attenuation_with_glucose}b), the depth of peak absorption should be approximately 3 mm $\times$ 4.5 dB/mm = 13.5 dB which is observed in measurement results (spectral response in Figure \ref{fig:Glucose_sensing_CPS_measurement_results}). We note that the simulated model only incorporates a single absorption peak at 1.42 THz as defined in Equation (\ref{eqn:dielectric}) for a uniform thin glucose layer covering the entire CPS. Also, we find our peak absorption frequency (1.42 THz) lies within the range (1.40 THz - 1.435 THz) found in other literature (see Table \ref{tab:glucose_sensing}). 


The frequency-dependent absorption observed in experimental results demonstrates the sensor functionality. It is important to recognize that this work is primarily a proof-of-concept and that future methods will be developed to improve sensitivity through the optimization of sample distribution methodology and circuit topology. We note that glucose shows broad absorption starting from 1.1 THz up to around 1.28 THz \cite{Lee_2015_Highly_gluceose_sensing_nano_antennas_nsrep,Song_2018_nsrep_absorption_1.42THz_solid_glucose, Liu2006_Dehydrated_Glucose_1.43_2.1_CPL}. This is observed in Figure \ref{fig:Glucose_sensing_CPS_measurement_results} as a minor depression before the main peak at 1.42 THz. Moreover, another THz absorption peak of glucose at 2.07 THz is observed in the measurement results and is expected to occur at 2.05 THz - 2.1 THz\cite{Liu2006_Dehydrated_Glucose_1.43_2.1_CPL,Song_2018_nsrep_absorption_1.42THz_solid_glucose, Jaber2024_Jean-michel_Menard_Nature_Communicaton, Lee_2015_Highly_gluceose_sensing_nano_antennas_nsrep}. We predict the attenuation of the dip at 2.07 THz relative to the attenuation at 1.42 THz using the data from \cite{Lee_2015_Highly_gluceose_sensing_nano_antennas_nsrep} which results in an expected attenuation of $\approx$ 5.8 dB at 2.08 THz. As shown in Figure \ref{fig:Glucose_sensing_CPS_measurement_results} the experimental results find a depth of absorption peak is $\approx$ 6.2 dB which is within reasonable agreement with the expected value.

It is important to consider practical aspects of sensors. The broadband nature of the received spectrum has several benefits which can reduce the probability of false positives and false negatives. First, the absorption at 1.42 THz can be measured relative to the adjacent region of the spectrum (i.e., 1.5 THz), but this method would be susceptible to false positives since other materials, such as sucrose, exhibit weak absorption characteristics near 1.4 THz. However, since the presented sensor is capable of resolving broad spectral information that corresponds to two signatures for glucose (1.42 THz and 2.07 THz) it becomes straightforward to reduce the probability of false positives by verifying that both signatures exist simultaneously with the appropriate relative magnitudes. Next, the sensor can exhibit false negatives if the glucose film is poorly distributed away from the CPS. Our proof-of-concept sensor is susceptible to this issue [see Figure \ref{fig:Membrane_Glucose_illustration}(b)] due to our current drop-coating method; however, we have begun investigating microfluidic methods to address this issue which are beyond the scope of this work.


\section{Conclusion}

An on-chip glucose sensor is demonstrated using guided waves at THz frequencies on a thin \ce{Si3N4} membrane. The sensor is based on the interaction between the propagating electromagnetic waves at THz frequencies and a thin layer of glucose placed in close proximity to a transmission line which results in frequency-dependent notches in the received spectrum. Our experiments reveal a resonance at 1.42 THz and 2.07 THz which aligns with findings in existing literature. Our approach stands out for its guided-wave nature and on-chip microfluidic integration potential. This work demonstrates pathways towards on-chip label-free sensing of glucose and other materials with applications such as the food industry, security, and healthcare.

\begin{acknowledgement}
 This work was supported by an NSERC Discovery Grant. The authors thank 4D LABS at Simon Fraser University for the fabrication of the CPS waveguides and the thin membrane, and also the Centre for Advanced Materials and Related Technology (CAMTEC) at the University of Victoria for providing Nanofab facilities for the fabrication of the PCS devices. We would like to thank Jean-Michel Ménard (and Ahmed Jaber) from the University of Ottawa for the inspiring discussions.
\end{acknowledgement}


\bibliography{Glucose_Sensing.bib}

\providecommand{\latin}[1]{#1}
\makeatletter
\providecommand{\doi}
  {\begingroup\let\do\@makeother\dospecials
  \catcode`\{=1 \catcode`\}=2 \doi@aux}
\providecommand{\doi@aux}[1]{\endgroup\texttt{#1}}
\makeatother
\providecommand*\mcitethebibliography{\thebibliography}
\csname @ifundefined\endcsname{endmcitethebibliography}
  {\let\endmcitethebibliography\endthebibliography}{}
\begin{mcitethebibliography}{35}
\providecommand*\natexlab[1]{#1}
\providecommand*\mciteSetBstSublistMode[1]{}
\providecommand*\mciteSetBstMaxWidthForm[2]{}
\providecommand*\mciteBstWouldAddEndPuncttrue
  {\def\EndOfBibitem{\unskip.}}
\providecommand*\mciteBstWouldAddEndPunctfalse
  {\let\EndOfBibitem\relax}
\providecommand*\mciteSetBstMidEndSepPunct[3]{}
\providecommand*\mciteSetBstSublistLabelBeginEnd[3]{}
\providecommand*\EndOfBibitem{}
\mciteSetBstSublistMode{f}
\mciteSetBstMaxWidthForm{subitem}{(\alph{mcitesubitemcount})}
\mciteSetBstSublistLabelBeginEnd
  {\mcitemaxwidthsubitemform\space}
  {\relax}
  {\relax}

\bibitem[Hassan \latin{et~al.}(2021)Hassan, Vyas, Grieve, and
  Bartolo]{Hassan2021_Glucose_Recent_advances_MDPI_Sensors}
Hassan,~M.~H.; Vyas,~C.; Grieve,~B.; Bartolo,~P. Recent Advances in Enzymatic
  and Non-Enzymatic Electrochemical Glucose Sensing. \emph{Sensors}
  \textbf{2021}, \emph{21}\relax
\mciteBstWouldAddEndPuncttrue
\mciteSetBstMidEndSepPunct{\mcitedefaultmidpunct}
{\mcitedefaultendpunct}{\mcitedefaultseppunct}\relax
\EndOfBibitem
\bibitem[Mozaffarian(2016)]{Mozaffarian2016_diabetes_obesity_circulation}
Mozaffarian,~D. Dietary and Policy Priorities for Cardiovascular Disease,
  Diabetes, and Obesity. \emph{Circulation} \textbf{2016}, \emph{133},
  187--225\relax
\mciteBstWouldAddEndPuncttrue
\mciteSetBstMidEndSepPunct{\mcitedefaultmidpunct}
{\mcitedefaultendpunct}{\mcitedefaultseppunct}\relax
\EndOfBibitem
\bibitem[Ventura \latin{et~al.}(2011)Ventura, Davis, and
  Goran]{ventura2011_obesity_diabetes_with_sugar}
Ventura,~E.~E.; Davis,~J.~N.; Goran,~M.~I. Sugar Content of Popular Sweetened
  Beverages Based on Objective Laboratory Analysis: Focus on Fructose Content.
  \emph{Obesity} \textbf{2011}, \emph{19}, 868--874\relax
\mciteBstWouldAddEndPuncttrue
\mciteSetBstMidEndSepPunct{\mcitedefaultmidpunct}
{\mcitedefaultendpunct}{\mcitedefaultseppunct}\relax
\EndOfBibitem
\bibitem[Oliver \latin{et~al.}(2009)Oliver, Toumazou, Cass, and
  Johnston]{Oliver2009_Review_Glucose_Sensors_diabetic_medicine}
Oliver,~N.~S.; Toumazou,~C.; Cass,~A. E.~G.; Johnston,~D.~G. Glucose sensors: a
  review of current and emerging technology. \emph{Diabetic Medicine}
  \textbf{2009}, \emph{26}, 197--210\relax
\mciteBstWouldAddEndPuncttrue
\mciteSetBstMidEndSepPunct{\mcitedefaultmidpunct}
{\mcitedefaultendpunct}{\mcitedefaultseppunct}\relax
\EndOfBibitem
\bibitem[Di~Filippo \latin{et~al.}(2023)Di~Filippo, Sunstrum, Khan, and
  Welsh]{Filipo2023_Review_Non-Invasive_Glucose_Sensing_MDPI_Sensors}
Di~Filippo,~D.; Sunstrum,~F.~N.; Khan,~J.~U.; Welsh,~A.~W. Non-Invasive Glucose
  Sensing Technologies and Products: A Comprehensive Review for Researchers and
  Clinicians. \emph{Sensors} \textbf{2023}, \emph{23}\relax
\mciteBstWouldAddEndPuncttrue
\mciteSetBstMidEndSepPunct{\mcitedefaultmidpunct}
{\mcitedefaultendpunct}{\mcitedefaultseppunct}\relax
\EndOfBibitem
\bibitem[Choi \latin{et~al.}(2022)Choi, Lee, Lee, Hong, and
  Kwon]{Choi_2022_natural_artificial_sweetener}
Choi,~Y.; Lee,~S.; Lee,~S.; Hong,~S.; Kwon,~H.~W. Bioelectronic Tongues
  Mimicking Insect Taste Systems for Real-Time Discrimination between Natural
  and Artificial Sweeteners. \emph{ACS Sensors} \textbf{2022}, \emph{7},
  3682--3691, PMID: 36455033\relax
\mciteBstWouldAddEndPuncttrue
\mciteSetBstMidEndSepPunct{\mcitedefaultmidpunct}
{\mcitedefaultendpunct}{\mcitedefaultseppunct}\relax
\EndOfBibitem
\bibitem[{American Chemical Society}(2012)]{D-glucose}
{American Chemical Society}, D-Glucose. 2012;
  \url{https://www.acs.org/molecule-of-the-week/archive/g/d-glucose.html}\relax
\mciteBstWouldAddEndPuncttrue
\mciteSetBstMidEndSepPunct{\mcitedefaultmidpunct}
{\mcitedefaultendpunct}{\mcitedefaultseppunct}\relax
\EndOfBibitem
\bibitem[Liu and Zhang(2006)Liu, and
  Zhang]{Liu2006_Dehydrated_Glucose_1.43_2.1_CPL}
Liu,~H.-B.; Zhang,~X.-C. Dehydration kinetics of D-glucose monohydrate studied
  using THz time-domain spectroscopy. \emph{Chemical Physics Letters}
  \textbf{2006}, \emph{429}, 229--233\relax
\mciteBstWouldAddEndPuncttrue
\mciteSetBstMidEndSepPunct{\mcitedefaultmidpunct}
{\mcitedefaultendpunct}{\mcitedefaultseppunct}\relax
\EndOfBibitem
\bibitem[Pickwell and Wallace(2006)Pickwell, and
  Wallace]{Pickwell_2006_THz_bio_applications}
Pickwell,~E.; Wallace,~V.~P. Biomedical applications of terahertz technology.
  \emph{Journal of Physics D: Applied Physics} \textbf{2006}, \emph{39},
  R301\relax
\mciteBstWouldAddEndPuncttrue
\mciteSetBstMidEndSepPunct{\mcitedefaultmidpunct}
{\mcitedefaultendpunct}{\mcitedefaultseppunct}\relax
\EndOfBibitem
\bibitem[Jacques(2013)]{Jacques_2013_Optical_properties_of_tissue_scattering}
Jacques,~S.~L. Optical properties of biological tissues: a review.
  \emph{Physics in Medicine and Biology} \textbf{2013}, \emph{58}, R37\relax
\mciteBstWouldAddEndPuncttrue
\mciteSetBstMidEndSepPunct{\mcitedefaultmidpunct}
{\mcitedefaultendpunct}{\mcitedefaultseppunct}\relax
\EndOfBibitem
\bibitem[Truong \latin{et~al.}(2022)Truong, Shahdadian, Kang, Wang, and
  Liu]{Truong2022_Influence_of_SNR_scatter_FOPT_}
Truong,~N. C.~D.; Shahdadian,~S.; Kang,~S.; Wang,~X.; Liu,~H. Influence of the
  signal-to-noise ratio on variance of chromophore concentration quantification
  in broadband near-infrared spectroscopy. \emph{Frontiers in Photonics}
  \textbf{2022}, \emph{3}\relax
\mciteBstWouldAddEndPuncttrue
\mciteSetBstMidEndSepPunct{\mcitedefaultmidpunct}
{\mcitedefaultendpunct}{\mcitedefaultseppunct}\relax
\EndOfBibitem
\bibitem[Kottmann \latin{et~al.}(2012)Kottmann, Rey, Luginb\"{u}hl, Reichmann,
  and Sigrist]{Kottmann2012_Glucose_epidermis_mid_IR_weak}
Kottmann,~J.; Rey,~J.~M.; Luginb\"{u}hl,~J.; Reichmann,~E.; Sigrist,~M.~W.
  Glucose sensing in human epidermis using mid-infrared photoacoustic
  detection. \emph{Biomed. Opt. Express} \textbf{2012}, \emph{3},
  667--680\relax
\mciteBstWouldAddEndPuncttrue
\mciteSetBstMidEndSepPunct{\mcitedefaultmidpunct}
{\mcitedefaultendpunct}{\mcitedefaultseppunct}\relax
\EndOfBibitem
\bibitem[Lee \latin{et~al.}(2015)Lee, Kang, Lee, Kim, Kim, Kim, Lee, Son, Park,
  and Seo]{Lee_2015_Highly_gluceose_sensing_nano_antennas_nsrep}
Lee,~D.; Kang,~J.; Lee,~J.~H.; Kim,~H.; Kim,~C.; Kim,~J.; Lee,~T.; Son,~J.;
  Park,~Q.; Seo,~M. Highly sensitive and selective sugar detection by terahertz
  nano-antennas. \emph{Scientific Reports} \textbf{2015}, \emph{5}\relax
\mciteBstWouldAddEndPuncttrue
\mciteSetBstMidEndSepPunct{\mcitedefaultmidpunct}
{\mcitedefaultendpunct}{\mcitedefaultseppunct}\relax
\EndOfBibitem
\bibitem[Vilagosh \latin{et~al.}(2019)Vilagosh, Lajevardipour, and
  Wood]{Vilagosh_2019_THz_penetration_tissue_0.2mm}
Vilagosh,~Z.; Lajevardipour,~A.; Wood,~A.~W. Computational phantom study of
  frozen melanoma imaging at 0.45 terahertz. \emph{Bioelectromagnetics}
  \textbf{2019}, \emph{40}, 118--127\relax
\mciteBstWouldAddEndPuncttrue
\mciteSetBstMidEndSepPunct{\mcitedefaultmidpunct}
{\mcitedefaultendpunct}{\mcitedefaultseppunct}\relax
\EndOfBibitem
\bibitem[Chen \latin{et~al.}(2018)Chen, Chen, Ma, \latin{et~al.}
  others]{Chen2018Quantify_glucose_THz_TDS}
Chen,~H.; Chen,~X.; Ma,~S., \latin{et~al.}  Quantify Glucose Level in Freshly
  Diabetic’s Blood by Terahertz Time-Domain Spectroscopy. \emph{J Infrared
  Milli Terahz Waves} \textbf{2018}, \emph{39}, 399--408\relax
\mciteBstWouldAddEndPuncttrue
\mciteSetBstMidEndSepPunct{\mcitedefaultmidpunct}
{\mcitedefaultendpunct}{\mcitedefaultseppunct}\relax
\EndOfBibitem
\bibitem[Lu \latin{et~al.}(2023)Lu, Xu, Luo, Li, Chang, Wei, and
  Cui]{Lu2023_WVA_THz_TDS_ACS_Photonics}
Lu,~X.; Xu,~L.; Luo,~L.; Li,~Z.; Chang,~T.; Wei,~D.; Cui,~H.-L. Weak Value
  Amplified Precision Terahertz Spectroscopic Detection of Solid and Liquid
  Glucose Samples. \emph{ACS Photonics} \textbf{2023}, \emph{10},
  3149--3160\relax
\mciteBstWouldAddEndPuncttrue
\mciteSetBstMidEndSepPunct{\mcitedefaultmidpunct}
{\mcitedefaultendpunct}{\mcitedefaultseppunct}\relax
\EndOfBibitem
\bibitem[Song \latin{et~al.}(2018)Song, Fan, Ding, Xu, Chen, and
  Wang]{Song_2018_nsrep_absorption_1.42THz_solid_glucose}
Song,~C.; Fan,~W.; Ding,~L.; Xu,~C.; Chen,~Z.; Wang,~K. Terahertz and infrared
  characteristic absorption spectra of aqueous glucose and fructose solutions.
  \emph{Scientific Reports} \textbf{2018}, \emph{8}\relax
\mciteBstWouldAddEndPuncttrue
\mciteSetBstMidEndSepPunct{\mcitedefaultmidpunct}
{\mcitedefaultendpunct}{\mcitedefaultseppunct}\relax
\EndOfBibitem
\bibitem[Jaber \latin{et~al.}(2024)Jaber, Reitz, Singh, Maleki, Xin, Sullivan,
  Dolgaleva, Boyd, Genes, and
  Ménard]{Jaber2024_Jean-michel_Menard_Nature_Communicaton}
Jaber,~A.; Reitz,~M.; Singh,~A.; Maleki,~A.; Xin,~Y.; Sullivan,~B.~T.;
  Dolgaleva,~K.; Boyd,~R.~W.; Genes,~C.; Ménard,~J. Hybrid architectures for
  terahertz molecular polaritonics. \emph{Nature Communications} \textbf{2024},
  \emph{15}\relax
\mciteBstWouldAddEndPuncttrue
\mciteSetBstMidEndSepPunct{\mcitedefaultmidpunct}
{\mcitedefaultendpunct}{\mcitedefaultseppunct}\relax
\EndOfBibitem
\bibitem[Lee \latin{et~al.}(2023)Lee, Cho, and
  Ok]{LEE2023_Imoroved_Modified_Lorentz_1.435_Spectrochimica}
Lee,~G.; Cho,~Y.; Ok,~G. Improved analysis of THz metamaterials for glucose
  sensing based on modified Lorentz dispersion model. \emph{Spectrochimica Acta
  Part A: Molecular and Biomolecular Spectroscopy} \textbf{2023}, \emph{293},
  122519\relax
\mciteBstWouldAddEndPuncttrue
\mciteSetBstMidEndSepPunct{\mcitedefaultmidpunct}
{\mcitedefaultendpunct}{\mcitedefaultseppunct}\relax
\EndOfBibitem
\bibitem[Huang \latin{et~al.}(2022)Huang, Shao, Wang, Su, and
  Zhang]{Huang_2022_glucose_spectrum_1.44THz}
Huang,~H.; Shao,~S.; Wang,~G.; Su,~B.; Zhang,~C. Terahertz spectral properties
  of glucose and two disaccharides in solid and liquid states. \emph{iScience}
  \textbf{2022}, \emph{25}, 104102\relax
\mciteBstWouldAddEndPuncttrue
\mciteSetBstMidEndSepPunct{\mcitedefaultmidpunct}
{\mcitedefaultendpunct}{\mcitedefaultseppunct}\relax
\EndOfBibitem
\bibitem[Beruete and Jáuregui-López(2019)Beruete, and
  Jáuregui-López]{Beruete_2019_AOM_Review_of_THz_sensing_metasurfaces}
Beruete,~M.; Jáuregui-López,~I. Terahertz sensing based on metasurfaces.
  \emph{Advanced Optical Materials} \textbf{2019}, \emph{8}\relax
\mciteBstWouldAddEndPuncttrue
\mciteSetBstMidEndSepPunct{\mcitedefaultmidpunct}
{\mcitedefaultendpunct}{\mcitedefaultseppunct}\relax
\EndOfBibitem
\bibitem[Yang \latin{et~al.}(2021)Yang, Qi, Li, Wu, Shi, Uqaili, and
  Tao]{Yang_2021_glucose_concentration_metasurface_400GHz}
Yang,~J.; Qi,~L.; Li,~B.; Wu,~L.; Shi,~D.; Uqaili,~J.~A.; Tao,~X. A terahertz
  metamaterial sensor used for distinguishing glucose concentration.
  \emph{Results in Physics} \textbf{2021}, \emph{26}, 104332\relax
\mciteBstWouldAddEndPuncttrue
\mciteSetBstMidEndSepPunct{\mcitedefaultmidpunct}
{\mcitedefaultendpunct}{\mcitedefaultseppunct}\relax
\EndOfBibitem
\bibitem[Serita \latin{et~al.}(2019)Serita, Murakami, Kawayama, and
  Tonouchi]{Serita_2019_MDPI_phptonic_THz_microfluidic_glucose_400GHz_meta}
Serita,~K.; Murakami,~H.; Kawayama,~I.; Tonouchi,~M. A terahertz-microfluidic
  chip with a few arrays of asymmetric meta-atoms for the ultra-trace sensing
  of solutions. \emph{Photonics} \textbf{2019}, \emph{6}, 12\relax
\mciteBstWouldAddEndPuncttrue
\mciteSetBstMidEndSepPunct{\mcitedefaultmidpunct}
{\mcitedefaultendpunct}{\mcitedefaultseppunct}\relax
\EndOfBibitem
\bibitem[Zeng \latin{et~al.}(2022)Zeng, Gong, Wang, Zhou, Zhang, Lan, Cong,
  Wang, Song, Zhao, Yang, and
  Mittleman]{Zeng2022_THz_PM_free-space_guided-wave}
Zeng,~H.; Gong,~S.; Wang,~L.; Zhou,~T.; Zhang,~Y.; Lan,~F.; Cong,~X.; Wang,~L.;
  Song,~T.; Zhao,~Y.; Yang,~Z.; Mittleman,~D.~M. A review of terahertz phase
  modulation from free space to guided wave integrated devices.
  \emph{Nanophotonics} \textbf{2022}, \emph{11}, 415--437\relax
\mciteBstWouldAddEndPuncttrue
\mciteSetBstMidEndSepPunct{\mcitedefaultmidpunct}
{\mcitedefaultendpunct}{\mcitedefaultseppunct}\relax
\EndOfBibitem
\bibitem[Singh \latin{et~al.}(2020)Singh, Tiwari, and
  Akhtar]{Singh_2020_SSPP_Sensor}
Singh,~S.~P.; Tiwari,~N.~K.; Akhtar,~M.~J. Spoof Surface Plasmon-Based Coplanar
  Waveguide Sensor for Dielectric Sensing Applications. \emph{IEEE Sensors
  Journal} \textbf{2020}, \emph{20}, 193--201\relax
\mciteBstWouldAddEndPuncttrue
\mciteSetBstMidEndSepPunct{\mcitedefaultmidpunct}
{\mcitedefaultendpunct}{\mcitedefaultseppunct}\relax
\EndOfBibitem
\bibitem[Cabello-Sánchez \latin{et~al.}(2021)Cabello-Sánchez, Drakinskiy,
  Stake, and Rodilla]{Calello_2021_On_chip_sensing_high_loss_liquids_CPW_TTST}
Cabello-Sánchez,~J.; Drakinskiy,~V.; Stake,~J.; Rodilla,~H. On-Chip
  Characterization of High-Loss Liquids Between 750 and 1100 GHz. \emph{IEEE
  Transactions on Terahertz Science and Technology} \textbf{2021}, \emph{11},
  113--116\relax
\mciteBstWouldAddEndPuncttrue
\mciteSetBstMidEndSepPunct{\mcitedefaultmidpunct}
{\mcitedefaultendpunct}{\mcitedefaultseppunct}\relax
\EndOfBibitem
\bibitem[Ohkubo \latin{et~al.}(2006)Ohkubo, Onuma, Kitagawa, and
  Kadoya]{Ohkubo2006_Microstrip_THz_liquid_sensing}
Ohkubo,~T.; Onuma,~M.; Kitagawa,~J.; Kadoya,~Y. {Micro-strip-line-based sensing
  chips for characterization of polar liquids in terahertz regime}.
  \emph{Applied Physics Letters} \textbf{2006}, \emph{88}, 212511\relax
\mciteBstWouldAddEndPuncttrue
\mciteSetBstMidEndSepPunct{\mcitedefaultmidpunct}
{\mcitedefaultendpunct}{\mcitedefaultseppunct}\relax
\EndOfBibitem
\bibitem[Cheng \latin{et~al.}(1994)Cheng, Whitaker, Weller, and
  Katehi]{cheng1994terahertz}
Cheng,~H.; Whitaker,~J.; Weller,~T.; Katehi,~L. Terahertz-bandwidth pulse
  propagation on a coplanar stripline fabricated on a thin membrane. \emph{IEEE
  microwave and guided wave letters} \textbf{1994}, \emph{4}, 89--91\relax
\mciteBstWouldAddEndPuncttrue
\mciteSetBstMidEndSepPunct{\mcitedefaultmidpunct}
{\mcitedefaultendpunct}{\mcitedefaultseppunct}\relax
\EndOfBibitem
\bibitem[Smith and Darcie(2019)Smith, and Darcie]{Levi_Smith_CPS_on_Si3N4_1st}
Smith,~L.; Darcie,~T. Demonstration of a low-distortion terahertz
  system-on-chip using a CPS waveguide on a thin membrane substrate.
  \emph{Optics Express} \textbf{2019}, \emph{27}, 13653--13663\relax
\mciteBstWouldAddEndPuncttrue
\mciteSetBstMidEndSepPunct{\mcitedefaultmidpunct}
{\mcitedefaultendpunct}{\mcitedefaultseppunct}\relax
\EndOfBibitem
\bibitem[Hu \latin{et~al.}(2016)Hu, Xu, Wen, Wang, Zhao, Zhang, Cumming, and
  Chen]{Hu_2016_metamaterial_absorber_glucose_ethanol}
Hu,~X.; Xu,~G.; Wen,~L.; Wang,~H.; Zhao,~Y.; Zhang,~Y.; Cumming,~D. R.~S.;
  Chen,~Q. Metamaterial absorber integrated microfluidic terahertz sensors.
  \emph{Laser and Photonics Reviews} \textbf{2016}, \emph{10}, 962--969\relax
\mciteBstWouldAddEndPuncttrue
\mciteSetBstMidEndSepPunct{\mcitedefaultmidpunct}
{\mcitedefaultendpunct}{\mcitedefaultseppunct}\relax
\EndOfBibitem
\bibitem[Cataldo \latin{et~al.}(2012)Cataldo, Beall, Cho, McAndrew, Niemack,
  and Wollack]{Cataldo_Silicon_nitride_properties_2012}
Cataldo,~G.; Beall,~J.~A.; Cho,~H.-M.; McAndrew,~B.; Niemack,~M.~D.;
  Wollack,~E.~J. Infrared dielectric properties of low-stress silicon nitride.
  \emph{Optics Letters} \textbf{2012}, \emph{37}, 4200--4202\relax
\mciteBstWouldAddEndPuncttrue
\mciteSetBstMidEndSepPunct{\mcitedefaultmidpunct}
{\mcitedefaultendpunct}{\mcitedefaultseppunct}\relax
\EndOfBibitem
\bibitem[Ríos \latin{et~al.}(2015)Ríos, Bikorimana, Ummy, Dorsinville, and
  Seo]{Rios2015_bowtie_PCA}
Ríos,~R. D.~V.; Bikorimana,~S.; Ummy,~M.~A.; Dorsinville,~R.; Seo,~S.-W. A
  bow-tie photoconductive antenna using a low-temperature-grown GaAs thin-film
  on a silicon substrate for terahertz wave generation and detection.
  \emph{Journal of Optics} \textbf{2015}, \emph{17}, 125802\relax
\mciteBstWouldAddEndPuncttrue
\mciteSetBstMidEndSepPunct{\mcitedefaultmidpunct}
{\mcitedefaultendpunct}{\mcitedefaultseppunct}\relax
\EndOfBibitem
\bibitem[Haghighat \latin{et~al.}(2024)Haghighat, Darcie, and
  Smith]{Haghighat2024_CPS_SSPP_SREP}
Haghighat,~M.; Darcie,~T.; Smith,~L. Demonstration of a terahertz
  coplanar-strip spoof-surface-plasmon-polariton low-pass filter.
  \emph{Scientific Reports} \textbf{2024}, \emph{14}\relax
\mciteBstWouldAddEndPuncttrue
\mciteSetBstMidEndSepPunct{\mcitedefaultmidpunct}
{\mcitedefaultendpunct}{\mcitedefaultseppunct}\relax
\EndOfBibitem
\bibitem[Jepsen \latin{et~al.}(2011)Jepsen, Cooke, and
  Koch]{Jepsen2011_THzSpec}
Jepsen,~P.; Cooke,~D.; Koch,~M. Terahertz spectroscopy and imaging – Modern
  techniques and applications. \emph{Laser \& Photonics Reviews} \textbf{2011},
  \emph{5}, 124--166\relax
\mciteBstWouldAddEndPuncttrue
\mciteSetBstMidEndSepPunct{\mcitedefaultmidpunct}
{\mcitedefaultendpunct}{\mcitedefaultseppunct}\relax
\EndOfBibitem
\end{mcitethebibliography}

\end{document}